\documentclass[12pt,preprint]{aastex}
\begin{document}

\bibliographystyle{apj}

%%%%%%%%%%%%%%%%%%%%%%%%%%%%%%%%%%%%%%%%%%%%%%%%%%%%%%%%%%%%%%%%%%%%

\title{Estimating Third-Order Moments for an Absorber Catalog}
\author{J. M. Loh}

\affil{Dept of Statistics, Columbia U, New York}
\email{meng@stat.columbia.edu}

\begin{abstract}
Due to recent availability of large surveys, there is renewed interest
in third-order correlation statistics. Measures of third-order
clustering are sensitive to the structure of filaments and voids in
the universe and are useful for studying large-scale
structure. Thus statistics of these third-order measures can be used to
test and constrain parameters in cosmological models.

Third-order measures such as the three-point correlation function are
now commonly estimated for galaxy surveys. Studies on third-order
clustering of absorption systems will complement these analyses.
We define a statistic, which we denote as $\mathcal{K}$,  that
measures third-order clustering of a 
dataset of point observations, and focus on the estimation of this
statistic for an absorber catalog. The statistic
$\mathcal{K}$ can be considered  
a third-order version of the second-order Ripley's $K$ function
and 
allows one to study the abundance of various
configurations of point triplets. In particular
configurations consisting of point triplets that lie close to a
straight line can be examined.

Studying third-order clustering of absorbers requires
consideration of the absorbers as a three-dimensional process,
observed on Quasi-stellar object (QSO) lines of
sight that extend radially in three-dimensional space from the
Earth. Since most of this three-dimensional space is not probed by the
lines of sight, edge corrections become important. We use an
analytical form of edge correction weights and
construct an estimator of the statistic $\mathcal{K}$ for use with an absorber catalog.
We show that with these
weights, ratio-unbiased estimates of $\mathcal{K}$ can be obtained.
Results from a simulation study also verify
unbiasedness and provide information on the decrease of standard
errors with increasing number of lines of sight.

\end{abstract}


\section{Introduction} \label{sect:intro}

The C \textsc{iv} and Mg \textsc{ii} Quasi-stellar object (QSO)
absorption systems or absorbers 
appear to trace the same structure as that of galaxies on very large
scales and have been shown to be effective probes of large-scale
structure in the universe \citep{crotts85a, crotts85b,
  tytler93}. See \citet{tripp05} for a discussion on the connection
between galaxies and QSO absorbers using observations at low redshift.

The clustering of such absorbers were studied in a series of
investigations \citep{vanden96, quashnock96, quashnock98} using an
extensive catalog of absorbers drawn from the
literature. Specifically, they performed a second-order correlation
analysis in one dimension, restricting to absorber pairs lying on the
same QSO lines of sight, and found clustering on very large scales, up
to 50 to 100 $h^{-1}$Mpc. This superclustering has also been found in
other studies, e.g.\ \citet{heisler89, dinshaw96}. This work on the
second-order clustering of absorbers complements the analyses of second-order
structure of other astronomical objects such as galaxies, quasars and
the cosmic microwave background.

In astronomy, a common measure of second-order structure is the
two-point correlation function $\xi$ \citep{peebles80, peebles93}, and this is the
function used in the above-mentioned studies. 
Another measure of second-order clustering is the reduced second
moment function, also called Ripley's $K$ function
\citep{ripley88, martinez02}. In three dimensions, the $K$ 
function is related to the two-point correlation function by
$$K(r) = 4\pi \int_0^r u^2[1+\xi(u)] \, du.$$
\citet{martinez98} applied the $K$ function to galaxy surveys, while
\citet{quashnock99, stein2001} used it to examine
clustering of the \citet{vanden96} C \textsc{iv} absorber
catalog.

\citet{loh01}
extended the work of \citet{quashnock99} in the study of second-order
clustering of the \citet{vanden96} absorber catalog 
by considering the absorbers as a process occurring in three
dimensions. By treating the absorbers as a three-dimensional process, 
absorber pairs that lie on different lines of sight were included in
estimates of the $K$ function. As a result, the
estimates obtained were shown to have dramatically smaller standard
errors than estimates obtained by only considering the absorbers as a
one-dimensional process on the lines of sight, when there is a large
enough number of lines of sight. 

More recently, there has been interest in higher-order clustering, 
in particular in third-order clustering, partly because of limitations
of restricting to second moments and partly because datasets are now
large enough for third-order statistics to be estimated. In
particular, the structure of filaments and voids that is present in
galaxy surveys is more readily, though still inadequately, described by
third-order statistics \citep{gaztanaga05a, sefusatti05}.
See \citet{jing98, gaztanaga05, nichol06, kulkarni07} for some examples
of the three-point correlation function \citep{fry80} applied to
galaxy surveys. The \citet{vanden96} absorber catalog, which has 276
lines of sight and 345 C \textsc{iv} absorbers, is too small for
investigating third-order structure, but the catalog being gathered by
the Sloan Digital Sky Survey \citep{york00} will have many more absorbers and
lines of sight, making a 
study of the third-order structure of absorbers feasible.

Here, we are concerned with estimating the third-order structure of an
absorber catalog. 
With estimates describing the third-order structure of an absorber catalog,
one can compare these estimates with corresponding estimates from
galaxy surveys. For example, one can study whether absorbers lie
along filaments like galaxies do.

In Section \ref{sect:thirdmoment} we will define a third-order version
of the $K$ function and relate it to the three-point correlation
function. We provide edge correction weights for its estimation for an
absorber catalog. We provide mathematical details in the Appendix
and results of a simulation study (Section \ref{sect:sim}) to show that
these weights do properly account for the edge effects.
Since these weights make use of the weights found in \citet{loh01}, we briefly
summarize their method of finding correction weights (Section
\ref{sect:absSecond}). Section \ref{sect:conclusion} contains a brief
summary and discussion of the application of this work for studying
galaxy clustering and large-scale structure.

\section{Estimation of $K$ for an absorber catalog}
\label{sect:absSecond}

Here we briefly describe the method of \citet{loh01} for finding
correction weights for estimating the $K$ function from an absorber
catalog.

\begin{figure}
\begin{center}
\plotone{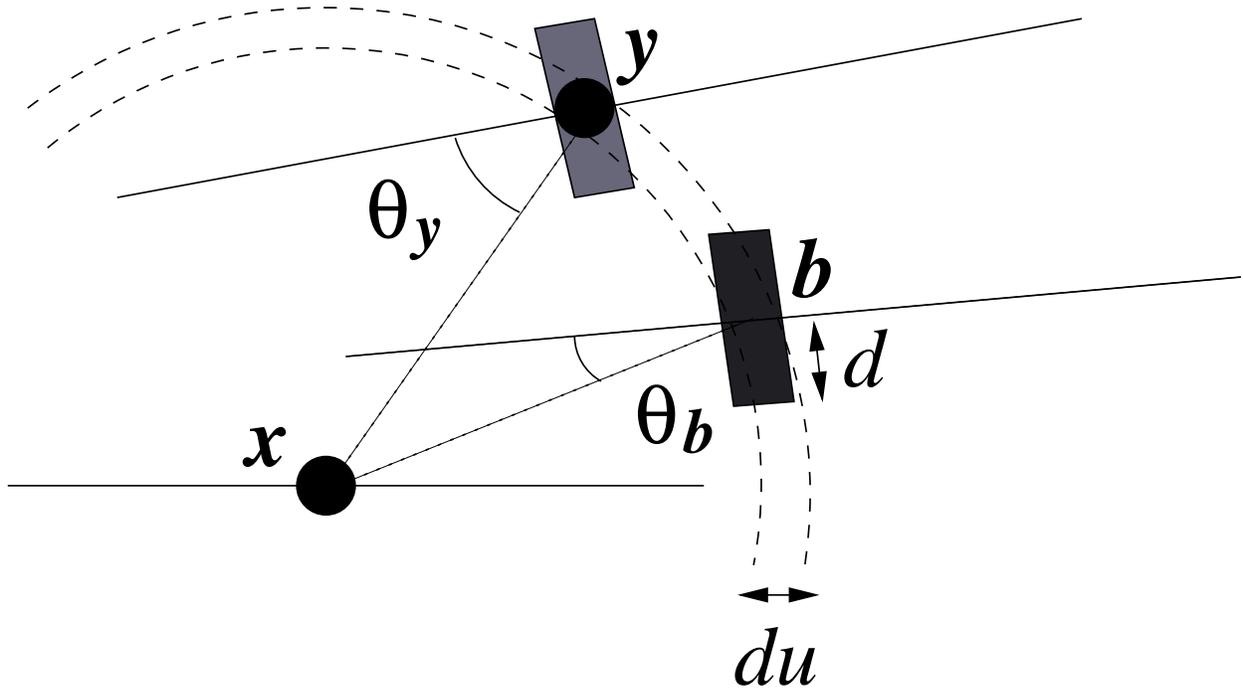}
\caption{A schematic diagram showing absorbers at $\mathbf{x}$ and
  $\mathbf{y}$. The solid lines represent lines of
  sight, and the dashed lines the shell centered at $\mathbf{x}$, with
  radius $|\mathbf{y}-\mathbf{x}|$ and thickness $du$. Each of the
  shaded rectangles represents cylinders in three-dimensional space
  and shows where an absorber center must lie
  in order to be detected at that particular location on a line of
  sight.}
\label{fig:pair}
\end{center}
\end{figure}

Figure \ref{fig:pair} is a schematic diagram that shows 
absorbers lying on some lines of sight. 
The solid lines represent lines of sight, and the solid circles, absorbers at $\mathbf{x}$ and $\mathbf{y}$. 
The dashed circles represent a shell centered at
$\mathbf{x}$ with radius $|\mathbf{y}-\mathbf{x}|$ and thickness $du$,
which we will denote by $\delta B_{du}(\mathbf{x},
|\mathbf{y}-\mathbf{x}|)$. This shell has volume $4\pi
|\mathbf{y}-\mathbf{x}|^2 du$. The point $\mathbf{b}$ represents an
intersection point of $\delta B_0(\mathbf{x},
|\mathbf{y}-\mathbf{x}|)$ and $L$, the set of
lines of sight. Note that with regards to notation, in this paper we will use
$\mathbf{x}, \mathbf{y}, 
\mathbf{z}$ to represent locations of absorbers on $L$ and $\mathbf{a},
\mathbf{b}, \mathbf{c}$ to refer to general locations on $L$ which may
or may not have absorbers present.

\clearpage
\clearpage

The two shaded rectangles on this shell represent cylinders in
three-dimensional space. If the center of the absorber $\mathbf{y}$
lies in a cylinder, it will be observed on the line of sight
that passes through the center of that cylinder. 
When the absorber at $\mathbf{x}$ is at the center, the contribution
of the absorber at $\mathbf{y}$ to the estimate of $K$ needs to be
corrected for boundary 
effects, i.e.\ for the points in the shell $\delta B_{du}(\mathbf{x},
|\mathbf{y}-\mathbf{x}|)$ not probed by the lines of
sight so that an absorber could be present there but not observable. Using
Ripley's method of edge correction, the 
weight is the reciprocal of the probability of detecting an absorber,
given that the absorber is present in the shell. This weight is
approximated by the ratio of  the volume of the shell $\delta B_{du}(\mathbf{x},
|\mathbf{y}-\mathbf{x}|)$ to the volume of the cylinders.
Specifically, the weight is given by
\begin{eqnarray}
w_\mathbf{x}(|\mathbf{y}-\mathbf{x}|) & = & \frac{4\pi
|\mathbf{y}-\mathbf{x}|^2 du}{ \pi d^2(du) \sum_{\mathbf{p}\in
  I_{\mathbf{x},\mathbf{y}}}(1/\cos
\theta_\mathbf{p})} =  \frac{4\pi
|\mathbf{y}-\mathbf{x}|^2 }{ \pi d^2 \sum_{\mathbf{p}\in
  I_{\mathbf{x},\mathbf{y}}}(1/\cos
\theta_\mathbf{p})},  \label{eqn:2ndorderwt} 
\end{eqnarray}
where $I_{\mathbf{x},\mathbf{y}}$ is the set of points in $\delta B_0(\mathbf{x},
|\mathbf{y}-\mathbf{x}|) \cap L$, and $\theta_\mathbf{p}$ is the angle
subtended by the line of sight that point $\mathbf{p}$ lies on, and
the line 
joining point $\mathbf{p}$ and absorber $\mathbf{x}$. In Figure
\ref{fig:pair}, $I_{\mathbf{x},\mathbf{y}}$ just consists of the
points $\mathbf{y}$ and $\mathbf{b}$. The angles $\theta_\mathbf{y}$
and  $\theta_\mathbf{b}$ are
also indicated in Figure \ref{fig:pair}. The variable
$d$ is the radius of an absorber in comoving units, and we assume it
to be unknown, but 
fixed. The value of $d$ does not need to be specified because it gets
cancelled away and does not appear in the estimator for $K$. For more
details, see
Section \ref{sect:thirdmoment} where this cancellation also occurs for
the estimator of the third-order statistic $\mathcal{K}$. 

With weights specified by (\ref{eqn:2ndorderwt}),
\citet{loh01} show that estimates for $\lambda^2 K$ are unbiased.

\section{Estimating third-order statistics for the absorber catalog}
\label{sect:thirdmoment} 

It is well-known that second-order statistics do not completely
describe the clustering properties of point processes. For example,
\citet{baddeley84} provide an example of a non-Poisson process with
the $K$ function identical to that of a homogeneous Poisson process.
Third- and higher-order statistics will allow a more detailed
study of clustering than just second-order statistics.

\citet{peebles75} defined a three-point correlation function $\zeta$ and
applied it to the Zwicky catalog. With three volume elements $dV_1,
dV_2$ and $dV_3$ that define a triangle with lengths $r_{12}, r_{23}$
and $r_{13}$, \citet{peebles75} wrote the probability of finding an
object in each of these elements as
\begin{eqnarray}
dP & = & \lambda^3 [1+\xi(r_{12})+\xi(r_{23})+\xi(r_{13})
+\zeta(r_{12},r_{23},r_{12})]dV_1dV_2dV_3, \label{eqn:threept}
\end{eqnarray}
where $\lambda$ is the intensity or number density of the point
process. 
Subsequent studies using the three-point correlation function
frequently used a different description of the configuration of
triplets, employing two distance measures and one angle measure:
$s=r_{12}, q=r_{23}/r_{12}$ and $\theta$, the angle between $r_{12}$
and $r_{23}$. See, for example, \citet{gaztanaga05,
  nichol06, kulkarni07}. Note that in the above notation the angle is
subtended at the 
second point. We will use the parametrization of two distances
and an angle in this work.

For the study of galaxy surveys, a related quantity $Q$, called
the reduced three-point correlation function is often used, where
\begin{eqnarray}
\zeta(s,q,\theta) = Q(s,q,\theta)\times [\xi(r_{12})\xi(r_{23}) +
\xi(r_{23})\xi(r_{13}) + \xi(r_{13})\xi(r_{12})]. \label{eqn:Q}
\end{eqnarray}
The hierarchical form of (\ref{eqn:Q}) was proposed in
\citet{peebles75} based on their analyses of the Lick and Zwicky
catalogs. It is an empirical form without theoretical support
\citep{jing98}, but has been found to hold in other studies e.g.\
\citet{szapudi01}. In analyses of the
2dFGRS and SDSS galaxy surveys, there appears to be variation of $Q$
with $\theta$ \citep[e.g.][]{gaztanaga05, nichol06}.

In the statistics literature, the quantity $dP$ in (\ref{eqn:threept})
above is more commonly expressed in terms of a function $g^{(3)}$:
\begin{eqnarray}
dP & = & \lambda^3g^{(3)}(r_{12},r_{23},\theta) dV_1dV_2dV_3, \label{eqn:g3}
\end{eqnarray}
where we have used two distances and an angle for the parameters of
$g^{(3)}$. 
\citet{moller98} refer to $\lambda^3 g^{(3)}$ as the third-product
density.
\citet{moller98} also
designed a third-order statistic to distinguish between certain
classes of point process models, while \citet{hanisch83} used a
third-order statistic to examine inner linearities in point patterns.
See also \citet{schladitz00}. These third-order statistics
are integrated versions of the third-product density, and
can be considered third-order versions of the second-order $K$ function. 
For our purposes, we define such a third-order function, which we
denote by $\mathcal{K}$:
\begin{eqnarray}
\mathcal{K}((0,R_1], (R_2,R_3], \Omega) & = &
\int_0^{R_1}\!\int_{R_2}^{R_3}\!\int_\Omega 4\pi r_{12}^2
r_{23}^2 g^{(3)}(r_{12},r_{23}, \theta) (2\pi \sin\theta)d\theta dr_{23}dr_{12}, \label{eqn:K}
\end{eqnarray}
where $(a,b]$ denotes an interval that includes $b$ but not $a$ and
$\Omega=[\alpha_1, \alpha_2], 0\le \alpha_1 < \alpha_2 \le \pi$ 
is a range of angles.
The third-order statistic of \citet{moller98} corresponds to
$\mathcal{K}$ with $R_2=0, R_3=R_1$ and $\Omega=[0,\pi]$.
The quantity $\mathcal{K}((0,R_1],
(R_2,R_3], \Omega)$ has an intuitive interpretation:\ given a randomly
chosen object at $\mathbf{x}$, $\lambda^2\mathcal{K}((0,R_1],
(R_2,R_3], \Omega)$ is the expected number of object pairs at $\mathbf{y}$ and
$\mathbf{z}$ such that $|\mathbf{y}-\mathbf{x}|\in (0, R_1],
|\mathbf{z}-\mathbf{x}| \in (R_2, R_3]$ and the angle subtended by
$\mathbf{y}$ and $\mathbf{z}$ at $\mathbf{x}$, $\angle \mathbf{yxz}$,
is in
$\Omega$. To relate to the notation in (\ref{eqn:K}), note that $r_{12}
= |\mathbf{y}-\mathbf{x}|$ and $r_{23} = 
|\mathbf{z}-\mathbf{x}|$, so that $\mathbf{x}$ is point 2, at which
the angle is subtended. Of
particular interest is the case 
when $\theta$ is close to 0 or $\pi$, since this describes the property of
finding triplets of points that lie close to a line. We will also be
interested in the variation of $\mathcal{K}$ with $\Omega$.

Both the three-point and reduced three-point correlation functions can
be obtained from $\mathcal{K}$.
Consider $\mathcal{K}((0,R_1], (0,R_3], \Omega)$ where $\Omega$ is a
small angle range, $(\theta-\delta\theta/2, \theta+\delta\theta/2)$,
say. Then from (\ref{eqn:K}) we have, 
\begin{eqnarray}
\frac{d^2\mathcal{K}}{dR_1dR_3} & = & 4\pi S(\Omega)R_1^2R_3^2
g^{(3)}(R_1, R_3, \theta), \label{eqn:dK}
\end{eqnarray}
where $S(\Omega)$ is the solid angle formed by the part of the unit
sphere that subtends an angle $\theta \in \Omega$ to the $x$-axis. 
Using (\ref{eqn:threept}), (\ref{eqn:Q}) and (\ref{eqn:g3}), $\zeta$ and
thus $Q$ can be expressed in terms of $g^{(3)}$. 
Therefore estimates for one
quantity can be converted to estimates for the other quantities.

There are different advantages to estimating $\mathcal{K}$ versus
$\zeta$. Since $\mathcal{K}$ is an integral quantity, it is often
smoother and thus its estimates might have better theoretical
properties. It also separates the choice of bin size from the edge
correction weights, so if a study of the effect of bin size or of
different edge correction methods is desired, it may be more
appropriate to use $\mathcal{K}$ \citep{stein2001}. On the other
hand, the hierarchical form of (\ref{eqn:Q}) is more simply expressed
using $\zeta$ and $Q$.
Having estimates of $\mathcal{K}, \zeta$ and $Q$ allows for more
flexibility in studying the clustering present in a dataset, so
rather than advocating for one statistic over another, we recommend
using all these statistics as tools for a detailed analysis.

When studying clustering of a point pattern observed in a finite
region, it is important to account for the boundary of the observation
region. If a point falls near a boundary, we do not get to
observe all its neighboring points. This is a particularly important
issue for
an absorber catalog, since only a small portion of the
three-dimensional space is probed by the lines of sight. In order to
obtain unbiased estimates, point pairs that are observed have
to be reweighted to account for the boundary effect. There are various
methods to do the edge correction. These can be numerical such as in
the estimators of the two-point 
correlation function introduced by \citet{davis83} and
\citet{hamilton93}, or analytical such as those introduced by
\citet{ripley88} and \citet{ohser83} for the $K$ function. See
\citet{kerscher2000} for a good review in the astronomy context.

\citet{loh01} found correction weights, based on the same correction
procedure suggested by \citet{ripley88}, for estimating the $K$
function for the \citet{vanden96} C \textsc{iv} absorber catalog. 
\citet{loh02} found expressions of the
correction weights based on Ohser's and Stoyan's correction
methods \citep{ohser81, ohser83}. Here, we obtain edge correction weights
for estimating the third-order moment function $\mathcal{K}$ using
Ripley's method.

Note that since estimating a third-order statistic involves counting
triplets of points, 
the more common analysis approach of treating the absorbers
as a one-dimensional process on the lines of sight cannot be used. The
absorbers have to be treated as a three-dimensional process, and the
edge effects caused by the large regions of unobserved space have to be
accounted for.

For fixed values of $R_1, R_2, R_3, \alpha_1$ and $\alpha_2$,
we estimate $\mathcal{K}$ by first estimating $\lambda^3A\mathcal{K}$ with
\begin{eqnarray}
 \sum_{\mathbf{x}\ne \mathbf{y}\ne \mathbf{z} \atop \mathbf{x},
   \mathbf{y}, \mathbf{z} \in L}
1_{(0,R_1]}(|\mathbf{y}-\mathbf{x}|)1_{(R_2,R_3]}(|\mathbf{z}-\mathbf{x}|)
1_\Omega(\angle\mathbf{yxz})
\omega_\mathbf{x}(|\mathbf{y}-\mathbf{x}|,|\mathbf{z}-\mathbf{x}|,\Omega)
V(|\mathbf{y}-\mathbf{x}|, |\mathbf{z}-\mathbf{x}|, \Omega), & &
\label{eqn:estimator} 
\end{eqnarray}
and then dividing the estimate $\widehat{\lambda^3A\mathcal{K}}$ by an
  estimator of $\lambda^3A$. Here, $L$ is the set of lines of sight,
  $A =\pi d^2|L|$ is the volume
probed by the lines of sight, $d$ is the constant radius of an
absorber in comoving units, $\angle\mathbf{yxz}$ 
is the angle subtended by $\mathbf{y}$ and $\mathbf{z}$ at
$\mathbf{x}$, and for any set $S$, $1_S(u )$ is an 
indicator function, equal to 1 if $u\in S$ and $0$ otherwise. 

We find that with
\begin{eqnarray}
\omega_\mathbf{x}(|\mathbf{y}-\mathbf{x}|,|\mathbf{z}-\mathbf{x}|, \Omega) &=
& \frac{4\pi S(\Omega)|\mathbf{y}-\mathbf{x}|^2 
  |\mathbf{z}-\mathbf{x}|^2(du)^2}{(\pi d^2)^2 (du)^2
\sum_{\mathbf{p}\in I_{\mathbf{x}, \mathbf{y}}}
\sum_{\mathbf{q}\in I_{\mathbf{x}, \mathbf{z}}}
 1_\Omega (\angle \mathbf{pxq})/(\cos \theta_\mathbf{p} \cos\theta_\mathbf{q})} \nonumber \\
& = & \frac{4\pi S(\Omega)|\mathbf{y}-\mathbf{x}|^2 
  |\mathbf{z}-\mathbf{x}|^2}{(\pi d^2)^2\
\sum_{\mathbf{p}\in I_{\mathbf{x}, \mathbf{y}}}
\sum_{\mathbf{q}\in I_{\mathbf{x}, \mathbf{z}}}
 1_\Omega (\angle \mathbf{pxq})/(\cos \theta_\mathbf{p} \cos\theta_\mathbf{q})},
\label{eqn:w} \\ 
V(|\mathbf{y}-\mathbf{x}|, |\mathbf{z}-\mathbf{x}|, \Omega) & = & \frac{A}{\pi
  d^2|L(|\mathbf{y}-\mathbf{x}|,|\mathbf{z}-\mathbf{x}|,\Omega)|} = \frac{|L|}{
  |L(|\mathbf{y}-\mathbf{x}|,|\mathbf{z}-\mathbf{x}|,\Omega)|},
\label{eqn:V}  
\end{eqnarray}
the estimator in (\ref{eqn:estimator}) is unbiased for
$\lambda^3A\mathcal{K}$. 
The quantity
$w_\mathbf{x}(|\mathbf{y}-\mathbf{x}|,|\mathbf{z}-\mathbf{x}|, \Omega)$ 
is the correction weight needed to account for the edge effects. It is
also called the local weight in 
\citet{kerscher2000}. The quantity $V(|\mathbf{y}-\mathbf{x}|,
|\mathbf{z}-\mathbf{x}|, \Omega)$ is sometimes 
referred to as Ohser's factor \citep{ohser83}, and makes the estimator
valid for longer distances. It is called a global weight in
\citet{kerscher2000}. For the rest of this section, we explain how the
expressions in (\ref{eqn:w}) and (\ref{eqn:V}) are obtained.

\begin{figure}
\begin{center}
\plotone{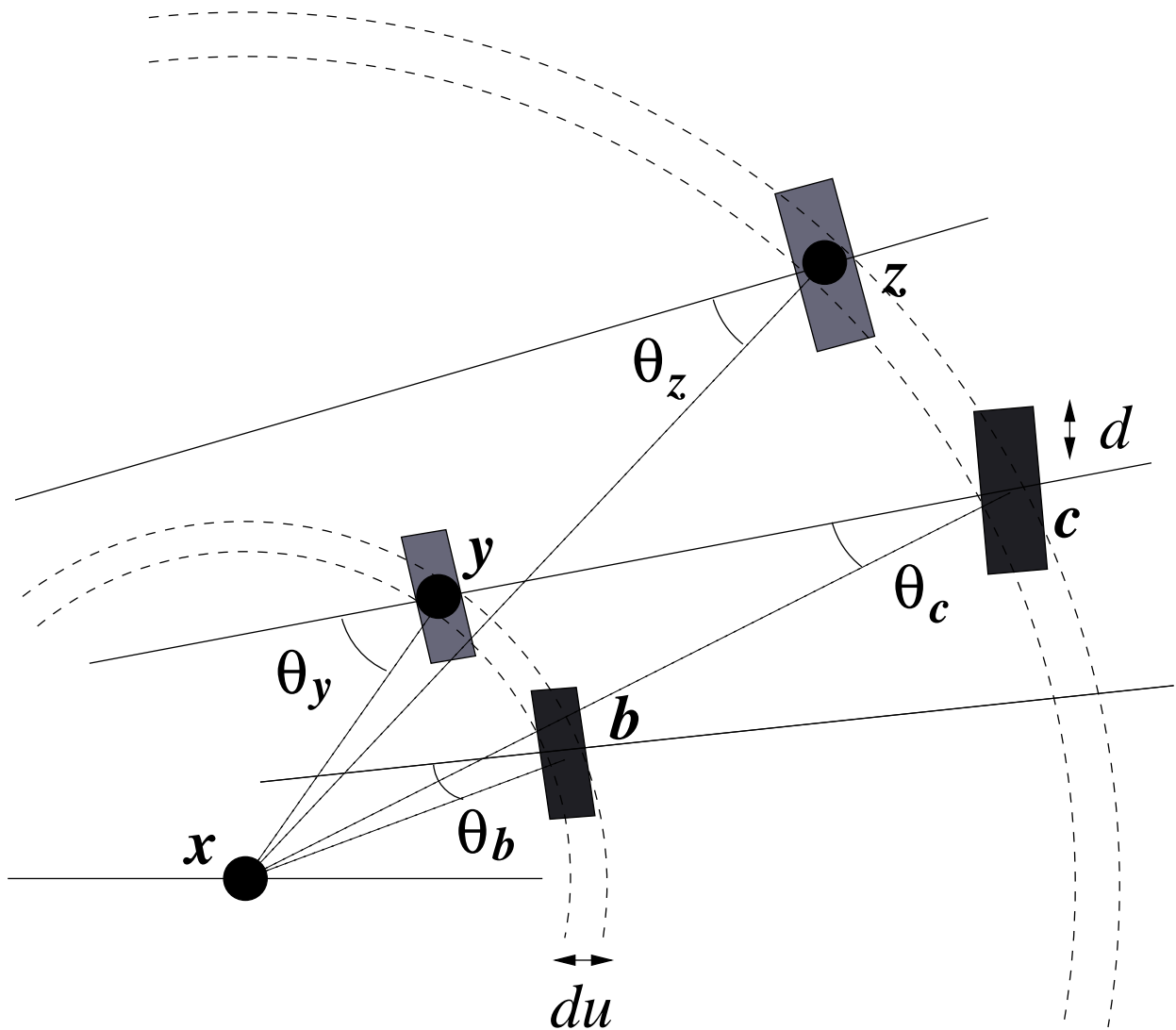}
\caption{A schematic diagram similar to Figure
  \ref{fig:pair}, with an additional absorber at $\mathbf{z}$, and a
  shell centered at $\mathbf{x}$, with 
  radius $|\mathbf{z}-\mathbf{x}|$ and thickness $du$. The triplet of
  absorbers $\mathbf{y},\mathbf{x}, \mathbf{z}$ is of the desired
  configuration. The triplet of points $\mathbf{b},
  \mathbf{x},\mathbf{c}$ corresponds to a set of locations in $L$ where
  a triplet of absorbers of the desired configuration could potentially
  have been observed.}
\label{fig:triplet}
\end{center}
\end{figure}

The denominator in the right-hand side of (\ref{eqn:w}) is
related to the weights found in \citet{loh01}, specifically, to the
denominator in the right-hand side of (\ref{eqn:2ndorderwt}).
In (\ref{eqn:2ndorderwt}), the denominator is the sum of the volumes of
the cylinders associated with the intersection of $\delta
B_{du}(\mathbf{x}, |\mathbf{y}-\mathbf{x}|)$ with the set of
lines of sight $L$, less a factor of $du$. These cylinders
are shown in Figure \ref{fig:pair} and are shown again in the lower
left portion of Figure \ref{fig:triplet}.

\clearpage
\clearpage

For (\ref{eqn:w}), we need to consider the intersections of $\delta
B_{du}(\mathbf{x}, |\mathbf{z}-\mathbf{x}|)$ with $L$ as well,
represented by the outer shell in Figure \ref{fig:triplet}. 
Like in (\ref{eqn:2ndorderwt}),
$I_{\mathbf{x},\mathbf{y}} = \delta
B_0(\mathbf{x},|\mathbf{y}-\mathbf{x}|) \cap L$ 
where $\delta B_0(\mathbf{x},|\mathbf{y}-\mathbf{x}|)$ is the sphere
centered at $\mathbf{x}$ with radius $|\mathbf{y}-\mathbf{x}|$. The
definition for $I_{\mathbf{x},\mathbf{z}}$ is similar. With respect to Figure
\ref{fig:triplet}, $I_{\mathbf{x}, \mathbf{y}}$ contains the locations
$\mathbf{b}$ and $\mathbf{y}$, while $I_{\mathbf{x}, \mathbf{z}}$
contains the locations 
$\mathbf{c}$ and $\mathbf{z}$.

To get the denominator on the right-hand side of (\ref{eqn:w}), we
consider pairs of cylinders, one on the outer shell and one on
the inner 
shell, i.e.\ a cylinder associated with a point $\mathbf{q}$ in
$I_{\mathbf{x},\mathbf{z}}$ and another associated with a point
$\mathbf{p}$ in 
$I_{\mathbf{x},\mathbf{y}}$. Each product of the volumes of these
pairs of cylinders, equal to $\pi d^2
(du) /\cos\theta_\mathbf{p} \times \pi d^2 (du)
/\cos\theta_\mathbf{q}$, is included in the 
sum only if the angle subtended at $\mathbf{x}$ by the centers of the
cylinder pair is in the range specified by $\Omega$, i.e.\ if $1_\Omega
(\mathbf{\angle pxq})=1$. In Figure
\ref{fig:triplet}, these pairs are highlighted by rectangles that are
similarly shaded. Note that the $(du)^2$ term cancels because there
is a corresponding term in the numerator of (\ref{eqn:w}). It is also
worth noting that the numerator of (\ref{eqn:w}) has a form
similar to
the right-hand side of (\ref{eqn:dK}).

There may be locations in $L$ that cannot be a possible location for
the absorber $\mathbf{x}$ of a triplet $\mathbf{y},\mathbf{x},
\mathbf{z}$ of the desired configuration. Which locations these are
depend on the actual 
positions and lengths of the 
lines of sight in $L$. The quantity $V$ in (\ref{eqn:V}) accounts for
this. Each location $\mathbf{a} \in L$ is in the set
$L(|\mathbf{y}-\mathbf{x}|, |\mathbf{z}-\mathbf{x}|, \Omega)$
if there are points $\mathbf{b},\mathbf{c} \in L$ such that
$|\mathbf{b}-\mathbf{a}| = |\mathbf{y}-\mathbf{x}|$,
$|\mathbf{c}-\mathbf{a}|=|\mathbf{z}-\mathbf{x}|$ and the angle
subtended by $\mathbf{b}$ and $\mathbf{c}$ at $\mathbf{a}$,
$\angle \mathbf{bac}$, is in $\Omega$, i.e.\ 
$L(|\mathbf{y}-\mathbf{x}|, |\mathbf{z}-\mathbf{x}|, \Omega)$ is just
the set $\{ \mathbf{a}\in L: \exists
\mathbf{b}, \mathbf{c}\in L \mbox{ with } |\mathbf{b}-\mathbf{a}| =
|\mathbf{y}-\mathbf{x}|,
|\mathbf{c}-\mathbf{a}|=|\mathbf{z}-\mathbf{x}|, \angle
\mathbf{bac}\in\Omega\}$. So, by definition, $\mathbf{x}$ of Figure
\ref{fig:triplet} has to be in 
$L(|\mathbf{y}-\mathbf{x}|, |\mathbf{z}-\mathbf{x}|, \Omega)$. 

To get an estimate of $\mathcal{K}$, we
divide the estimator (\ref{eqn:estimator}) by an estimate of $\lambda^3A$,
e.g.\ $(N^3/A^3)A = N^3/(\pi d^2)^2|L|^2$. Thus, although the
expression for $\omega_\mathbf{x}$ includes a $(\pi d^2)^2$ term,
the value of $d$ need not be specified when estimating $\mathcal{K}$
since it gets cancelled away by the same term in the estimate of $\lambda^3A$.

The proof of unbiasedness is provided in the Appendix. Note that it is
the estimator of $\lambda^3A\mathcal{K}$ that is unbiased. The
estimate $\hat{\mathcal{K}}$ that is obtained by dividing by an
estimate of $\lambda^3A$ may be slightly biased. Such a property is
called ratio-unbiasedness, and is a feature of estimators of the
second-order $K$ function as well.

\section{Simulation Study}
\label{sect:sim}

\begin{figure}
\begin{center}
\plotone{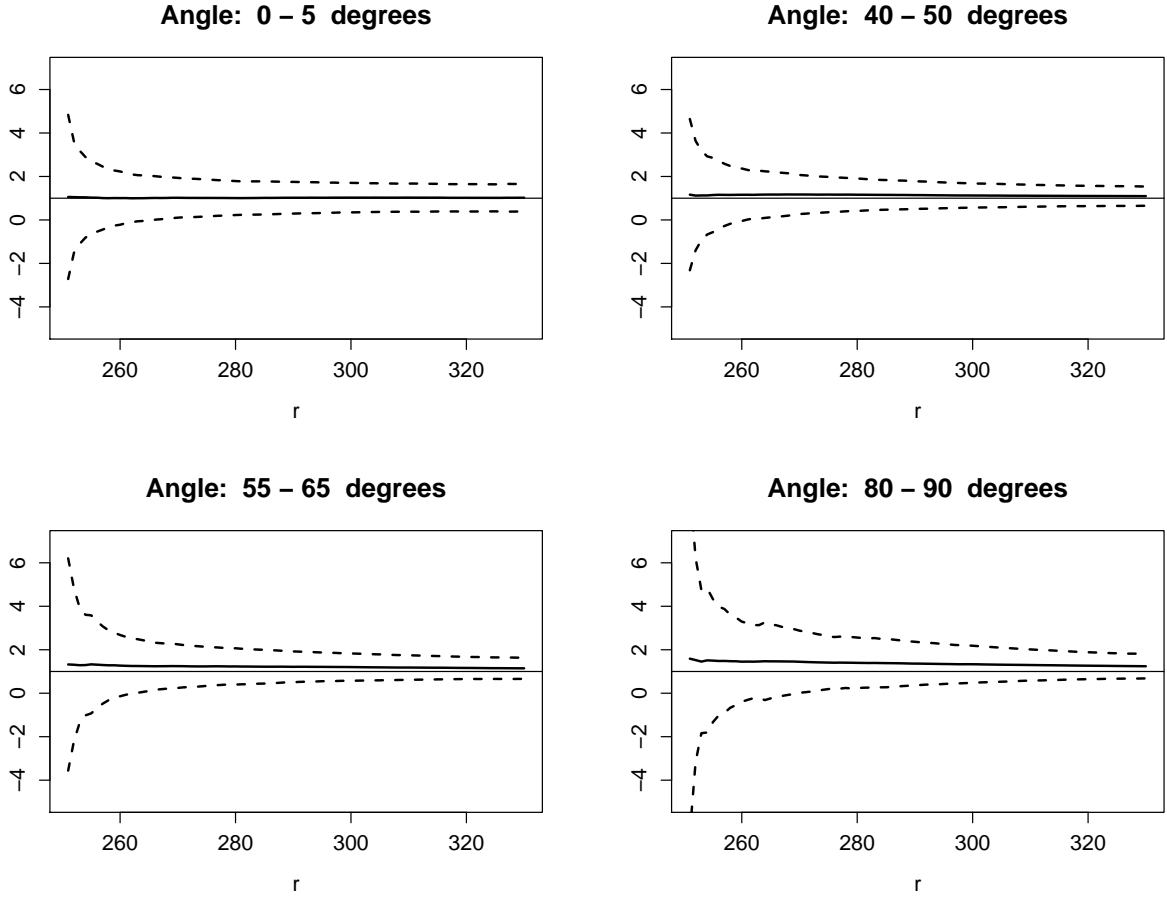}
\caption{Plots of the mean, over 1000 simulated realizations, of the
  ratio of $\hat{\mathcal{K}}((0,50], 
  (250,r], \Omega)$ 
  to the expected Poisson value as a function of $r$ from 250 to
  330 $h^{-1}$ Mpc. The $\Omega$ for each plot is specified at the top of
  plot. The dashed lines refer to pointwise errors that are twice the
  standard deviation obtained from the simulations.}
\label{fig:K3toP3}
\end{center}
\end{figure}

\begin{figure}
\begin{center}
\plotone{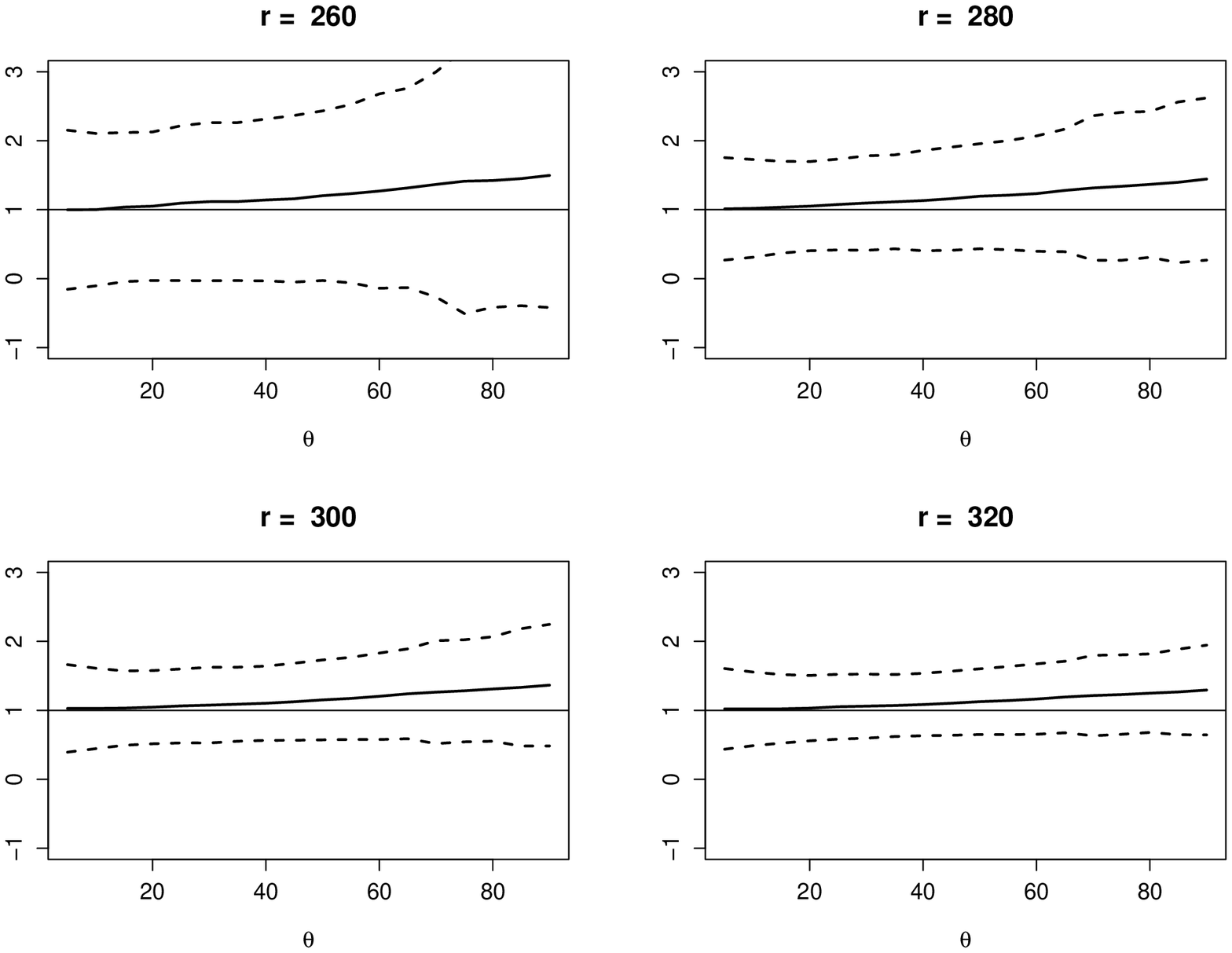}
\caption{Plots of the mean, over 1000 simulated realizations, of the
  ratio of $\hat{\mathcal{K}}((0,50], 
  (250,r], \theta)$ 
  to the expected Poisson value as a function of $\theta$ from 0 to
  $90^\circ$, using angular bins of $10^\circ$. The value of $r$ for each plot is specified at the top
  of plot. The dashed lines refer to pointwise errors that are twice
  the standard deviation obtained from the simulations.}
\label{fig:K3toP3byangle}
\end{center}
\end{figure}

We ran a simulation study to explore the performance of the estimator
given in (\ref{eqn:estimator}), with weights given in (\ref{eqn:w})
and (\ref{eqn:V}). Note that distances referred to here are comoving distances.
We first generated a set of 1000 lines of sight in a region similar to
that to be probed by the QSO lines of sight of the SDSS Catalog: a
cone with half-angle of $45^\circ$ with Earth at its tip, bounded by
comoving distance $2000  < r < 3300 h^{-1}$ Mpc from
Earth. This range of distances corresponds to the comoving distances
probed by QSO lines of sight for Mg \textsc{ii} and C \textsc{iv}
absorbers under the Einstein-de Sitter cosmology.
A thousand realizations of a Poisson point process are then
simulated on to these lines of sight, with density equal to that found
in the 
\citet{vanden96} catalog, 0.004 per $h^{-1}$ Mpc. We chose Poisson
processes since the 
theoretical value of $\mathcal{K}$ is known for the Poisson model:
$P=\mathcal{K}_{\mbox{\small Poi}}((0,R_1],(R_2,R_3],\Omega) = 4\pi S(\Omega) 
(R_3^3-R_2^3)R_1^3/9$. For each realization, we
estimate the third-order function $\mathcal{K}$. We then find the mean
and variance of these estimates, and compare it with the theoretical
Poisson value $P$. The results are shown in Figures \ref{fig:K3toP3} and
\ref{fig:K3toP3byangle}.

Figure \ref{fig:K3toP3} shows the ratio of the mean estimates of
$\mathcal{K}((0,50], (250, r], \Omega)$ to the expected Poisson value,
for $\Omega=(0^\circ,5^\circ)$ (top
left), $\Omega=(40^\circ,50^\circ)$ (top right), 
$\Omega=(55^\circ,60^\circ)$ (bottom left) and 
$\Omega=(80^\circ,90^\circ)$ (bottom right), plotted as a function of
$r$, for $250 <r\le 330$ $h^{-1}$ Mpc (solid lines). 
The dashed lines show the pointwise error, equal to
two times the standard deviation of the 1000 estimates. Notice that in
each case, the true value of 1 lies within this band. Furthermore the
mean estimated value is very close to 1 for the smaller angle ranges,
with a slight bias appearing with angles close to $90^\circ$. 
We believe this is because the edge correction approximation becomes
less accurate at angles close to $90^\circ$.

Figure \ref{fig:K3toP3byangle} shows plots of the same ratio
as a function of $\theta$, the midpoint of $\Omega$, from $0^\circ$ to
$90^\circ$, for values of $r$ fixed at 260, 280, 300 and 320 $h^{-1}$ Mpc. The
angular bin size used is $10^\circ$. Again, we find that the pointwise
confidence band contains the true value 1, with the mean
value also close to 1. These plots show the bias
appearing as the angle is increased to $90^\circ$, with this bias
becoming slightly less as the range $(250,r]$ increases.

\clearpage
\clearpage

\clearpage
\clearpage

We chose the distance 50 $h^{-1}$ Mpc for $R_1$ since it corresponds roughly to
the scale of superclustering that has been detected
\citep{quashnock96}. With the values of $r_{23}\in (250, 330]$ that we used, the ratio
$r_{23}/r_{12}$ is then about 5 or 6, close to the values considered in
e.g.\ \citet{kulkarni07}, although the values of $r_{12}$ considered
there are much smaller. 
We also considered values of
10 to 40 $h^{-1}$  Mpc for $R_1$. The results are qualitatively
similar to the results for $R_1=50 h^{-1}$ Mpc, except with slightly larger
standard errors. 

We also performed a simulation study using 10,000
lines of sight in the same region. The corresponding plots are similar
to those of Figures \ref{fig:K3toP3} and \ref{fig:K3toP3byangle}, but
with standard errors smaller by a factor of about 10 to 15, i.e. roughly of the
order of the increase in the number of lines of sight. For $\Omega =
(0^\circ,5^\circ)$, standard errors dropped only by about a factor of
5, however. This approximate relation between standard errors and the
number of lines of sight is similar to that found by \citet{loh01}.

\section{Discussion and Conclusion}
\label{sect:conclusion}

Measures of the third-order clustering of galaxy surveys and the
cosmic microwave background are useful for the additional information
they provide over measures of second-order clustering. In particular, 
the filamentary structure that has been found in galaxy data is more readily
described by third- and higher-order measures of clustering.
Recently, due to the availability of larger datasets and advances in
computing, such study of higher-order clustering has been the subject
of active research.

It will be desirable to study the third-order clustering of absorption
systems. Absorbers are often detected at extreme
comoving distances from the Earth. Since absorbers are believed to be
due to gas clouds near galaxies, an analysis of the third-order
clustering of absorbers can serve as a complementary analysis to that
of large galaxy surveys, enabling comparison of the local filamentary
structure to that of the early universe.
Studying the third-order structure of absorbers might also provide greater
understanding of their nature. Finally, absorption systems consist of
non-luminous matter. Understanding the clustering of absorbers can
yield insight into the link between luminous and non-luminous matter
in the universe.

In this paper, we define a third-order moment function $\mathcal{K}$
that is an integrated version of the three-point correlation function,
much like the relation between the second-order Ripley's $K$ function
and the two-point correlation function. We provide expressions for the
weights necessary to correct for the boundary effects so that this
function can be estimated for an absorber catalog. Our
simulation study shows that the estimator gives correct results (i.e.\
including correctly accounting for the boundary effects), at least for
the theoretically simple Poisson process. 

Studies on large-scale structure with galaxy surveys have shown the
existence of structures of the order of 
100 $h^{-1}$ Mpc in size \citep{kirshner81, geller89,
  costa94}. In analyses of second-order clustering of the Las Campanas
and SDSS surveys, \citet{landy96} and \citet{eisenstein05} respectively found peaks on
scales of around 100 $h^{-1}$ Mpc. \citet{quashnock96} also found
evidence of superclustering on these scales in their analysis of C
\textsc{iv} absorption systems. \citet{loh01} also found evidence of
clustering up to 100 $h^{-1}$ Mpc and possibly beyond.
Studies on galaxy clustering have focused on smaller
scales. \citet{gaztanaga05, nichol06} and \citet{kulkarni07} followed
the example of \citet{jing98}, using $r_{12}$ from 1 to 10 $h^{-1}$
Mpc and $r_{23}/r_{12}$ between 1 and 4, and studied the variation in
the reduced three-point correlation function $Q$ with angle.

The choice of $R_1, R_2, R_3$ and $\Omega$ for $\mathcal{K}$ would thus
depend on the aim of the analysis. For comparisons with the findings
of e.g. \citet{jing98}, the focus will be to study the variation of
$\mathcal{K}$ with $\Omega$, with the distance measures close
to the values used there. For studies on superclustering and
large-scale structure, comoving distances of 100 $h^{-1}$ Mpc and
beyond for one or both of $r_{12}$ and $r_{23}$ will be of
interest. An initial study will probably use $\Omega=[0,\pi]$,
studying clustering at various distances, followed by more detailed
analyses with smaller angular ranges.

We are not aware of any other work 
on estimating third-order clustering specifically
for absorber catalogs. Unfortunately, we do not have a large
enough catalog of absorbers to obtain meaningful estimates of the
third-order function.
With the much larger absorber catalog that is being collected by the
Sloan Digital Sky Survey, a detailed study of the third-order
clustering of absorbers will become feasible. 
From our simulation studies,
we found that the standard errors of estimates of $\mathcal{K}$ for an
absorber catalog scale roughly on the order of the reciprocal of the
number of lines of sight. This agrees with the findings of \citet{loh01} for
standard errors of estimates of the $K$ function for absorber
catalogs. Thus we expect that with the SDSS absorber catalog with
approximately 50,000 lines of sight, the standard errors of the
estimates of $\mathcal{K}$ will be roughly a factor of 50 smaller than
the standard errors found in our simulation study with 1000 lines of
sight. The actual increase in precision for a particular absorber
catalog will of course depend on factors such as the actual 
spatial locations of the lines of sight and the density of the
observed absorbers.

\acknowledgments

This research is supported in part by
National Science Foundation award AST-0507687.

\appendix
\section{Appendix}
\label{sect:appendix}

Here, we prove that the estimator (\ref{eqn:estimator}) with weights
$\omega_\mathbf{x}$ and $V$ given by (\ref{eqn:w}) and (\ref{eqn:V}) is
unbiased. Let $L$ represent the lines of sight, $\delta B_\Delta
(\mathbf{x}, h)$ denote a shell with center $\mathbf{x}$, radius $h$
and thickness $\Delta$. We write $(h,\gamma)$ for the polar
coordinates of vector $\mathbf{h}$, $|\cdot |$ for Euclidean distance,
area or volume depending on the context, $A=\pi d^2 |L|$ for the
volume probed by the lines of sight and $1_L(\mathbf{x})$ for the
indicator function, with $1_L(\mathbf{x})=1$ if $\mathbf{x}\in L$ and
0 otherwise.

Write $f(\mathbf{x},\mathbf{y},\mathbf{z}) =
1_L(\mathbf{x})1_L(\mathbf{y})1_L(\mathbf{z})1_{(0,R_1]}
(|\mathbf{y}-\mathbf{x}|)1_{(R_2,R_3]}(|\mathbf{z}-\mathbf{x}|)
1_\Omega(\angle \mathbf{yxz})
\omega_\mathbf{x}(|\mathbf{y}-\mathbf{x}|,|\mathbf{z}-\mathbf{x}|,\Omega)
V(|\mathbf{y}-\mathbf{x}|, |\mathbf{z}-\mathbf{x}|, \Omega))$, 
where $\Omega = [\alpha_1, \alpha_2]$ represents the range of angles between
$\alpha_1$ and $\alpha_2$. Then the estimator in (\ref{eqn:estimator})
is $\sum_{\mathbf{x}\ne \mathbf{y}\ne \mathbf{z}}
f(\mathbf{x},\mathbf{y},\mathbf{z})$, with
\begin{eqnarray*}
E\left(\sum_{\mathbf{x}\ne \mathbf{y}\ne \mathbf{z}} f(x,y,z)\right) &= & \lambda^3 \int\!\!\!\int\!\!\!\int
f(\mathbf{x}, \mathbf{x}+\mathbf{h}, \mathbf{x}+\mathbf{k})
g^{(3)}(\mathbf{x}, \mathbf{x}+\mathbf{h}, \mathbf{x}+\mathbf{k})\,
d\mathbf{h}\, d\mathbf{k}\, d\mathbf{x} \\
& = & \lambda^3\int_{\mathbb{R}^3}\int_0^{R_1}\!\!\! \int_{R_2}^{R_3}
\!\!\!\int_{\delta B_0(\mathbf{0}, k)} \int_{\delta B_0(\mathbf{0},
  h)}  1_L(\mathbf{x}) 
  1_L(\mathbf{x}+(h,\gamma))1_L(\mathbf{x}+(k,\beta)) \\
& & {}\times 1_\Omega(\gamma-\beta ) \omega_{\mathbf{x}}(h,k,\Omega)
  V(h,k,\Omega) h^2k^2 g^{(3)}((h,\gamma), (k,\beta) )\, d\gamma \,
  d\beta \, dh\, dk \, d\mathbf{x}.
\end{eqnarray*}
In the first equality above, we have expressed $g^{(3)}$ in terms of
three vector quantities $\mathbf{x}, \mathbf{x}+\mathbf{h}$ and
$\mathbf{x}+\mathbf{k}$. If stationarity is assumed,
the specification of $\mathbf{x}$ in $g^{(3)}$ is redundant. Thus we
have removed the dependence on $\mathbf{x}$ in $g^{(3)}$ in the next
line. We have also expressed $\mathbf{h}$ and $\mathbf{k}$ in polar
coordinates. 
Now, with the further assumption of isotropy,
$g^{(3)}(\mathbf{h},\mathbf{k})$ depends only on the direction of
$\mathbf{h}$ relative to $\mathbf{k}$ (or vice versa). This simplifies
the expression above, so that
\begin{eqnarray*}
E\left(\sum_{\mathbf{x}\ne \mathbf{y}\ne\mathbf{z}}
f(\mathbf{x},\mathbf{y},\mathbf{z})\right)  
& = & \lambda^3 \int_{\mathbb{R}^3}\int_0^{R_1}\!\!\! \int_{R_2}^{R_3}
\!\!\! V(h,k,\Omega) 1_L(\mathbf{x}) \omega_\mathbf{x}(h,k,\Omega)
\\
& & {}\times \int_{\delta B_0(\mathbf{0},k)} 1_L(\mathbf{x}+(k,\beta))
k^2 \\
& & {} \quad \times 
\left[ \int_{\delta B_0(\mathbf{0},h)} 1_{\Omega}
  (\alpha)1_L(\mathbf{x}+(h,\beta+\alpha)) h^2 g^{(3)}(h, k,
  \alpha)\, d\alpha \right] \, d\beta\, dh\, dk\, d\mathbf{x},
\end{eqnarray*}
where $\alpha$ denotes the angle on the sphere relative to $(k,\beta)$.
Under the assumption that $g^{(3)}$ is slowly varying over
$\Omega$, the expression in the square bracket above is equal to
$$1_{L(h,\beta+\Omega)}(\mathbf{x})g^{(3)}(h, k, \Omega)
\sum_{\mathbf{p} \in \delta B_0(\mathbf{x},h) \cap L}
1_{\Omega} (\angle \mathbf{px}(\mathbf{x}+\mathbf{k})) \frac{\pi d^2}{\cos
  \theta_\mathbf{p}},$$
so that 
\begin{eqnarray*}
E\left(\sum_{\mathbf{x}\ne \mathbf{y}\ne \mathbf{z}}
f(\mathbf{x},\mathbf{y},\mathbf{z})\right) 
& = & \lambda^3A \int_{\mathbb{R}^3}\int_0^{R_1}\!\!\! \int_{R_2}^{R_3}
\!\!\! \frac{1_L(\mathbf{x})}{\pi d^2|L(h,k,\Omega)|}
\omega_\mathbf{x}(h,k,\Omega) g^{(3)}(h,k, \Omega)\\ 
& & {}\times \left[ \int_{\delta B_0(\mathbf{0}, k)}
  1_{L(h,\beta+\Omega)}(\mathbf{x}) 1_L(\mathbf{x}+(k,\beta)) \right.
\\
& & {} \quad \times \left. \left\{
    \sum_{\mathbf{p} \in \delta B_0(\mathbf{x},h) \cap L} 1_\Omega
    (\angle \mathbf{px}(\mathbf{x}+\mathbf{k})) \frac{\pi d^2}{\cos
      \theta_\mathbf{p}} 
  \right\} k^2\, d\beta  \right]\, dh\, dk\, d\mathbf{x} \\
& = & \lambda^3A \int_{\mathbb{R}^3}\int_0^{R_1}\!\!\! \int_{R_2}^{R_3}
\!\!\! \frac{\omega_{\mathbf{x}} (h,k,\Omega)}{\pi d^2|L(h,k,\Omega)|}
1_{L(h,k,\Omega)}(\mathbf{x}) g^{(3)}(h,k, \Omega) \\
& & {}\times \left( \sum_{\mathbf{p} \in \delta B_0(\mathbf{x},h) \cap
    L} \sum_{\mathbf{q} \in \delta B_0(\mathbf{x},k) \cap L} 1_\Omega
  (\angle\mathbf{pxq})\frac{\pi d^2}{\cos \theta_\mathbf{p}}\frac{\pi
    d^2}{\cos \theta_\mathbf{q}} \right)\, dh\, dk\,
d\mathbf{x}.  
\end{eqnarray*}
The above expression in the round brackets is the denominator of
$\omega_\mathbf{x}(h,k,\Omega)$. Further simplification yields
\begin{eqnarray*}
E\left(\sum_{\mathbf{x}\ne \mathbf{y}\ne \mathbf{z}}
f(\mathbf{x},\mathbf{y},\mathbf{z})\right) 
& = & \lambda^3A \int_{\mathbb{R}^3}\int_0^{R_1}\!\!\! \int_{R_2}^{R_3}
\!\!\! \frac{1_{L(h,k,\Omega)}(\mathbf{x})}{\pi d^2 |L(h,k,\Omega)|} 4\pi
h^2k^2 S(\Omega)  g^{(3)}(h,k, \Omega)\, dh\, dk\, d\mathbf{x} \\
& = & \lambda^3A \int_0^{R_1}\!\!\! \int_{R_2}^{R_3}
\!\!\! 4\pi h^2k^2S(\Omega) g^{(3)}(h,k,\Omega)\, dh\, dk \\
&= & \lambda^3A\mathcal{K}((0,R_1], (R_2,R_3], \Omega).
\end{eqnarray*}


\begin{thebibliography}{42}
\expandafter\ifx\csname natexlab\endcsname\relax\def\natexlab#1{#1}\fi

\bibitem[{Baddeley \& Silverman(1984)}]{baddeley84}
Baddeley, A.~J., \& Silverman, B.~W. 1984, Biometrics, 40, 1089

\bibitem[{Crotts(1985)}]{crotts85a}
Crotts, A. P.~S. 1985, ApJ, 298, 732

\bibitem[{Crotts {et~al.}(1985)Crotts, Melott, York, \& Fry}]{crotts85b}
Crotts, A. P.~S., Melott, A.~L., York, D.~G., \& Fry, J.~N. 1985, Phys.
  Lett. B, 155, 251

\bibitem[{{da Costa} {et~al.}(1994){da Costa}, {et~al.}}]{costa94}
{da Costa}, L.~N., {et~al.} 1994, ApJ, 424, L1

\bibitem[{Davis \& Peebles(1983)}]{davis83}
Davis, M., \& Peebles, P. J.~E. 1983, ApJ, 267, 465

\bibitem[{Dinshaw \& Impey(1996)}]{dinshaw96}
Dinshaw, N., \& Impey, C.~D. 1996, ApJ, 458, 73

\bibitem[{Eisenstein {et~al.}(2005)Eisenstein, Zehavi, Hogg, \&
  Scoccimarro}]{eisenstein05}
Eisenstein, D.~J., Zehavi, I., Hogg, D.~W., \& Scoccimarro, R. 2005,
  ApJ, 633, 560

\bibitem[{Fry \& Peebles(1980)}]{fry80}
Fry, J.~N., \& Peebles, P. J.~E. 1980, ApJ, 238, 785

\bibitem[{Gazta{\~n}aga {et~al.}(2005)Gazta{\~n}aga, Norberg, Baugh, \&
  Croton}]{gaztanaga05}
Gazta{\~n}aga, E., Norberg, P., Baugh, C.~M., \& Croton, D.~J. 2005, MNRAS, 364, 620

\bibitem[{Gazta{\~n}aga \& Scoccimarro(2005)}]{gaztanaga05a}
Gazta{\~n}aga, E., \& Scoccimarro, R. 2005, MNRAS, 361, 824

\bibitem[{Geller \& Huchra(1989)}]{geller89}
Geller, M.~J., \& Huchra, J.~P. 1989, Science, 246, 897

\bibitem[{Hamilton(1993)}]{hamilton93}
Hamilton, A. J.~S. 1993, ApJ, 417, 19

\bibitem[{Hanisch(1983)}]{hanisch83}
Hanisch, K.-H. 1983, Math. Oper. Ser. Statist., 14, 421

\bibitem[{Heisler {et~al.}(1989)Heisler, Hogan, \& White}]{heisler89}
Heisler, J., Hogan, C.~J., \& White, S. D.~M. 1989, ApJ, 347,
  52

\bibitem[{Jing \& B\"{o}rner(1998)}]{jing98}
Jing, Y.~P., \& B\"{o}rner, G. 1998, ApJ, 503, 37

\bibitem[{Kerscher {et~al.}(2000)Kerscher, Szapudi, \& Szalay}]{kerscher2000}
Kerscher, M., Szapudi, I., \& Szalay, A.~S. 2000, ApJ, 535, L13

\bibitem[{Kirshner {et~al.}(1981)Kirshner, Oemler, Schechter, \&
  Shectman}]{kirshner81}
Kirshner, R.~P., Oemler, A., Schechter, P.~L., \& Shectman, S.~A. 1981,
  ApJ, 248, L57

\bibitem[{Kulkarni {et~al.}(2007)Kulkarni, Nichol, Sheth, Seo, Eisenstein, \&
  Gray}]{kulkarni07}
Kulkarni, G.~V., Nichol, R.~C., Sheth, R.~K., Seo, H.-J., Eisenstein, D.~J., \&
  Gray, A. 2007, MNRAS, 378, 1196

\bibitem[{Landy {et~al.}(1996)Landy, Schectman, Lin, Kirshner, Oemler, \&
  Tucker}]{landy96}
Landy, S.~D., Schectman, S.~A., Lin, H., Kirshner, R.~P., Oemler, A.~A., \&
  Tucker, D. 1996, ApJ, 456, L1

\bibitem[{Loh {et~al.}(2001)Loh, Quashnock, \& Stein}]{loh01}
Loh, J.~M., Quashnock, J.~M., \& Stein, M.~L. 2001, ApJ, 560,
  606

\bibitem[{Loh {et~al.}(2003)Loh, Stein, \& Quashnock}]{loh02}
Loh, J.~M., Stein, M.~L., \& Quashnock, J.~M. 2003, J. Am. Stat. Assoc.,
  98, 522

\bibitem[{Mart\'{\i}nez {et~al.}(1998)Mart\'{\i}nez, Pons-Border\'{\i}a,
  Moyeed, \& Graham}]{martinez98}
Mart\'{\i}nez, V.~J., Pons-Border\'{\i}a, M.-J., Moyeed, R.~A., \& Graham,
  M.~J. 1998, MNRAS, 298, 1212

\bibitem[{Mart\'{\i}nez \& Saar(2002)}]{martinez02}
Mart\'{\i}nez, V.~J., \& Saar, E. 2002, Statistics of the Galaxy Distribution
  (Boca Raton: Chapman and Hall/CRC)

\bibitem[{M{\o}ller {et~al.}(1998)M{\o}ller, Syversveen, \&
  Waagepetersen}]{moller98}
M{\o}ller, J., Syversveen, A.~R., \& Waagepetersen, R.~P. 1998, Scand.
  J. Stat., 25, 451

\bibitem[{Nichol {et~al.}(2006)Nichol, {et~al.}}]{nichol06}
Nichol, R.~C., {et~al.} 2006, MNRAS, 368, 1507

\bibitem[{Ohser(1983)}]{ohser83}
Ohser, J. 1983, Math. Oper. Ser. Statist., 14, 63

\bibitem[{Ohser \& Stoyan(1981)}]{ohser81}
Ohser, J., \& Stoyan, D. 1981, Biometric J., 23, 523

\bibitem[{Peebles(1980)}]{peebles80}
Peebles, P. J.~E. 1980, The Large-Scale Structure of the Universe (New Jersey:
  Princeton University Press)

\bibitem[{Peebles(1993)}]{peebles93}
---. 1993, Principles of Physical Cosmology (New Jersey: Princeton University
  Press)

\bibitem[{Peebles \& Groth(1975)}]{peebles75}
Peebles, P. J.~E., \& Groth, E.~J. 1975, ApJ, 196, 1

\bibitem[{Quashnock \& Stein(1999)}]{quashnock99}
Quashnock, J.~M., \& Stein, M.~L. 1999, ApJ, 515, 506

\bibitem[{Quashnock \& {Vanden Berk}(1998)}]{quashnock98}
Quashnock, J.~M., \& {Vanden Berk}, D.~E. 1998, ApJ, 500, 28

\bibitem[{Quashnock {et~al.}(1996)Quashnock, {Vanden Berk}, \&
  York}]{quashnock96}
Quashnock, J.~M., {Vanden Berk}, D.~E., \& York, D.~G. 1996, ApJ, 472, L69

\bibitem[{Ripley(1988)}]{ripley88}
Ripley, B.~D. 1988, Statistical Inference for Spatial Processes (New York:
  Wiley)

\bibitem[{Schladitz \& Baddeley(2000)}]{schladitz00}
Schladitz, K., \& Baddeley, A.~J. 2000, Scand. J. Stat., 27,
  657

\bibitem[{Sefusatti \& Scoccimarro(2005)}]{sefusatti05}
Sefusatti, E., \& Scoccimarro, R. 2005, Phys. Rev. D, 71, 063001

\bibitem[{Stein {et~al.}(2000)Stein, Quashnock, \& Loh}]{stein2001}
Stein, M.~L., Quashnock, J.~M., \& Loh, J.~M. 2000, Ann. Stat., 28,
  1503

\bibitem[{Szapudi {et~al.}(2001)Szapudi, Postman, Lauer, \&
  Oegerie}]{szapudi01}
Szapudi, I., Postman, M., Lauer, T., \& Oegerie, W. 2001, ApJ, 548, 114

\bibitem[{Tripp \& Bowen(2005)}]{tripp05}
Tripp, T.~M., \& Bowen, D.~V. 2005, in Probing Galaxies through Quasar
  Absorption Lines: Proc. IAU 199

\bibitem[{Tytler {et~al.}(1993)Tytler, Sandoval, \& Fan}]{tytler93}
Tytler, D., Sandoval, J., \& Fan, X.-M. 1993, ApJ, 405, 57

\bibitem[{{Vanden Berk} {et~al.}(1996){Vanden Berk}, Quashnock, York, \&
  Yanny}]{vanden96}
{Vanden Berk}, D.~E., Quashnock, J.~M., York, D.~G., \& Yanny, B. 1996,
  ApJ, 469, 78

\bibitem[{York {et~al.}(2000)York, {et~al.}}]{york00}
York, D.~G., {et~al.} 2000, AJ,  120, 1579

\end{thebibliography}
\end{document}